\documentclass{article}
\usepackage{arxiv}
\usepackage[utf8]{inputenc} 
\usepackage[T1]{fontenc}    
\usepackage{hyperref}       
\usepackage{multicol,multirow}
\usepackage{url}            
\usepackage{booktabs}       
\usepackage{amsmath,amssymb,amsfonts}
\usepackage{nicefrac}       
\usepackage{microtype}      
\usepackage{lipsum}		
\usepackage{graphicx}
\usepackage{natbib}
\usepackage{doi}
\usepackage{tabularx}

\title{Disentangling regional impacts of joint teleconnections using causal representation learning}

\author{ Fiona R. Spuler \\
	Department of Meteorology\\
	University of Reading, Reading, UK\\
	\texttt{f.r.spuler@pgr.reading.ac.uk} \\
	\And
	Marlene Kretschmer \\
	Leipzig Institute for Meteorology\\
	University of Leipzig,\\
        Leipzig, Germany \\
  	\And
	Magdalena Alonso Balmaseda \\
	European Centre for Medium-Range \\ Weather Forecasts\\
        Reading, UK \\
    \And
    Masilin Gudoshava \\
	IGAD Climate Prediction and Applications Centre\\
    Nairobi, Kenya\\
   	\And
	Theodore G. Shepherd \\
	Department of Meteorology,\\
        University of Reading \\
        Reading, UK \\
}

\date{}

\renewcommand{\headeright}{Spuler et al. 2026}
\renewcommand{\undertitle}{}
\renewcommand{\shorttitle}{}

\hypersetup{
pdftitle={DAG-VAE preprint},
pdfsubject={q-bio.NC, q-bio.QM},
pdfauthor={Fiona Spuler},
pdfkeywords={causal representation learning, teleconnections, seasonal predictability, variational autoencoders, regional impacts},
}

\begin{document}

\maketitle


\textbf{Abstract:} Understanding teleconnections of large-scale modes of climate variability is relevant for seasonal predictability and support a dynamical understanding of climatic changes. While numerical model experiments are the most common approach for investigating counterfactual climate responses, their conclusions are subject to model biases. Data-driven approaches offer a complementary perspective. Deep learning can extract reduced-dimensional patterns but usually lacks causal interpretability, while causal methods can disentangle signals in the presence of confounding yet are typically based on simple indices. Treating dimensionality reduction and causal inference separately thereby risks losing the teleconnection signal of interest. This paper introduces DAG-VAE, a causal representation learning approach that embeds a physics-informed directed acyclic graph in the latent space of a variational autoencoder. Combining deep learning with causal inference, the method jointly learns nonlinear reduced representations of large-scale modes of variability and their causal interactions. We apply DAG-VAE to disentangle the influences of the Pacific and Indian Oceans on the short rains over the Greater Horn of Africa. Trained on seasonal hindcasts, the method identifies dynamically meaningful representations and recovers spatial response patterns consistent with SST-replacement experiments. Trained on reanalysis data, DAG-VAE identifies a different response pattern to direct influence of the tropical Pacific, highlighting potential model biases and the value of DAG-VAE as a complementary, data-driven approach for estimating spatial causal response patterns from observations. Finally, we demonstrate the ability of the method to generate data-driven counterfactuals of extreme short rain seasons, with potential applications for forecast-based early action and scenario planning.

\textbf{Significance statement:} Investigating how large-scale climate variability affects regional rainfall is important for improving seasonal forecasts and the dynamical understanding of future climate change. Numerical model experiments are commonly used to infer causal teleconnections, but are subject to model biases. Data-driven methods can be trained on observations, yet they often struggle to disentangle causal pathways, especially when spatial patterns are complex and nonlinear. Here, we address this problem by developing a data-driven method that jointly learns low-dimensional representations of climate variability and their causal links, combining causal inference with deep learning. Applied to the Greater Horn of Africa, the method successfully disentangles teleconnection patterns from the Indian and Pacific Oceans and enables the generation counterfactual simulations of plausible extreme short-rain seasons.

\section{Introduction}\label{introduction}

Large-scale modes of variability in the climate system and their teleconnections provide important sources of predictability at subseasonal to seasonal timescales and a dynamical foundation for understanding near-term to end-of-century climate change \citep{mariotti_windows_2020, line_modulation_2024, mindlin_explaining_2025}. However, developing a robust dynamical understanding of teleconnections and disentangling their joint effects on surface-level extremes remains challenging due to model biases in representing teleconnections \citep{stan_advances_2022, gler_systematic_2025} and the relatively short observational record, background-state dependence and non-stationarity of teleconnections, as well as the difficulty in identifying causal effects in high-dimensional spatiotemporal data \citep{runge_detecting_2019, saggioro_reconstructing_2020, wainwright_eastern_2019}.\\

Over the Greater Horn of Africa (GHA), the seasonal movement of the Intertropical Convergence Zone modulates rainfall variability, leading to a spatially varying bimodal climatology of ``long rains'' (March-May) and ``short rains'' (October-December) \citep{palmer_drivers_2023, nicholson_climate_2017}. The short rain season, while more predictable, exhibits strong interannual variability and has been associated with both droughts and heavy rainfall, causing severe impacts on lives and livelihoods in recent years \citep{gudoshava_advances_2024, nana_diverse_2025}. \\

At seasonal timescales, the intensity and predictability of the short rains is influenced by teleconnections from the tropical Pacific and Indian Ocean, in particular the El Niño Southern Oscillation (ENSO) and Indian Ocean Dipole (IOD) through the modulation of the Indian Ocean Walker circulation \citep{gudoshava_atmospheric_2022, black_observational_2003, black_relationship_2005, kolstad_lagged_2022, palmer_drivers_2023, tefera_seasonal_2025}. ENSO is a coupled ocean–atmosphere phenomenon in the tropical Pacific and a leading mode of interannual climate variability globally \citep{timmermann_ninosouthern_2018}, while the IOD describes a dipole pattern of sea surface temperature anomalies in the Indian Ocean which peaks between September and November \citep{saji_dipole_1999, webster_coupled_1999}. El Niño and positive IOD conditions typically lead to anomalous subsidence over the eastern Indian Ocean, a reversal of the climatological westerlies, and enhanced convection and precipitation over the GHA, with the reverse effect observed during La Niña and negative IOD conditions \citep{palmer_drivers_2023, zheng_indian_2025}. \\

Although ENSO is considered to be one of the processes influencing the IOD, the IOD can also occur independently \citep{wainwright_extreme_2021, wang_indian_2024, cai_pantropical_2019, sun_triggering_2015}. Therefore, disentangling whether the seasonal teleconnection from the two ocean basins is primarily mediated via the Indian Ocean, or whether the tropical Pacific can exert a direct influence, independent of the IOD, remains challenging. Understanding this direct influence has important consequences for applications such as early warnings and anticipatory action. For instance, during the 2015 El Niño, flood preparedness action was taken in the humanitarian sector in Somalia and Kenya, based on past anomalies associated with strong El Niño events \citep{tozier_de_la_poterie_understanding_2018}. However, the short rains that year were less intense and more spatially variable than forecast, in part due to weaker than forecast sea surface temperature (SST) anomalies over the Indian Ocean \citep{macleod_moderate_2019}. This highlights the importance of better understanding whether the precipitation response to El Niño is entirely mediated by the in-phase occurrence of a positive IOD, or whether a direct effect from the tropical Pacific exists. \\

The causal effects of remote SST anomalies on regional precipitation have traditionally been explored using model experiments, such as SST-replacement experiments \citep{richter_tropical_2025}. Early studies found that the region-wide wetting or drying anomaly over the GHA is primarily mediated via the Indian Ocean, rather than via a direct pathway from the tropical Pacific \citep{latif_role_1999, goddard_importance_1999}. However, SST replacement experiments conducted by \cite{macleod_causal_2021} showed that while the primary pathway leading to an overall wetting or drying response is indeed mediated by the Indian Ocean, there is an additional direct pathway from the tropical Pacific to the GHA short rains that exhibits a different spatial precipitation response pattern: instead of an overall wetting or drying over the region, the authors argue that in the model considered, the direct effect of ENSO leads to a weaker dipole pattern response with a wetting off the coast of the GHA and a drying further inland in response to El Niño, and the opposite for La Niña. \\

While model experiments enable conditioning on specific states of a remote driver, their findings are necessarily subject to model errors and biases. Seasonal forecast systems as well as climate projections tend to underestimate the positive skewness of both ENSO and the IOD, exhibit mean state biases in the IOD that influence climatological precipitation over East Africa, and show an emerging El Niño-like response to anthropogenic emissions not found in observations \citep{gler_systematic_2025, wang_indian_2024, timmermann_ninosouthern_2018, beverley_climate_2024, hirons_impact_2018}. The relationship between ENSO and IOD has also been hypothesised to be too strong in some forecasting systems \citep{macleod_moderate_2019}. Furthermore, depending on the research question, the unphysical constraint imposed by a hard SST boundary condition that does not allow for atmosphere-ocean feedbacks, leading to unrealistic heat sources or sinks, poses limitations on the findings \citep{oreilly_challenges_2023, barsugli_basic_1998}. Numerical model experiments are also computationally expensive, which prevents their use in real time, limiting the ability to understand and communicate the predicted seasonal forecast anomaly effectively, in turn affecting decision-making based on early warnings.\\

These limitations call into question the extent to which model-based causal inferences hold in the observed climate system and motivate the development of complementary data-driven approaches. Recent advances in causality research have spurred the development of new data-driven methods for investigating teleconnections in a causal framework \citep{kretschmer_using_2016, kretschmer_role_2020, runge_inferring_2019, saggioro_probabilistic_2024, di_capua_dominant_2020}, based on the mathematical formalisation and representation of causal mechanisms as directed acyclic graphs \citep{pearl_causal_2009, peters_elements_2017}. These methods are based on either discovering causal graphs from time series data by ruling out non-causal relations using constraint-based \citep{runge_quantifying_2014, runge_detecting_2019} or score-based methods \citep{peters_elements_2017}, or assuming the causal graph based on physical hypotheses \citep{kretschmer_quantifying_2021}. \\

However, existing causal approaches often rely on predefined indices as representations of the modes of variability studied - such as the Niño3.4 for ENSO or the Dipole Mode Index (DMI) for the IOD \citep{kolstad_lagged_2022}. While such indices can provide useful standardization for the scientific community, they risk averaging out the signal of interest and obscuring physically meaningful patterns. In particular, both ENSO and IOD exhibit what is known as diversity, namely that different SST anomaly patterns with different surface-level impacts, drivers, and responses to global warming can be associated with the same index \citep{capotondi_understanding_2015, guo_three_2015, fischer_two_2005, du_new_2013, johnson_applied_2013, guillaume-castel_enso_nodate, zhang_physics-informed_2024, fang_nonlinear_2024}. These diverse SST patterns have also been shown to influence their teleconnections to the short rains \citep{wang_indian_2024, tozuka_anomalous_2016, liu_why_2017, macleod_causal_2021}. On the other hand, common dimensionality reduction approaches for representing large-scale modes, such as Empirical Orthogonal Function (EOF) analysis (also called principal component analysis; PCA), assume linearity, are not necessarily physically interpretable \citep{dommenget_cautionary_2002}, and treat dimensionality reduction separately from causal inference \citep{saji_dipole_1999, wang_indian_2024}. In the application studied in this paper, the precipitation dipole identified by \cite{macleod_causal_2021} could not be identified using the first principal component or regional average precipitation over the GHA used for example in \cite{kolstad_drivers_2021}. In addition, \cite{kolstad_beyond_2024} investigate the drivers of different spatial precipitation patterns over the GHA and highlight, along with \cite{nicholson_climate_2017}, the need to go beyond the regional average of precipitation for better understanding drivers and predicting impacts. \\

Deep learning approaches, such as graph neural networks, on the other hand, also learn nonlinear latent representations. However, these are optimized for prediction rather than representation of the underlying dynamical processes and are not necessarily physically or causally interpretable. The emerging field of causal representation learning aims to address this limitation by jointly learning reduced representations of high-dimensional data and their causal structure \citep{scholkopf_toward_2021}. While the majority of papers in this field have focused on established applications in computer science such as computer vision, a few recent papers have proposed methods to investigate causal links in the climate system \citep{boussard_towards_2023, wang_uncovering_2024, hickman_causal_2025}. These methods primarily work towards climate model emulation and make general assumptions about the structure of the latent space, such as the single parent structure assumed in \cite{boussard_towards_2023}, but do not easily incorporate existing knowledge about dynamical processes in the model structure.\\

In this work, building on these recent advances in causal representation learning, we propose a new data-driven method to jointly identify dimensionality-reduced representations of climatic processes alongside the causal effects of interacting teleconnections. Our proposed method, DAG-VAE, embeds a physics-informed directed acyclic graph (DAG) in the latent space of a variational autoencoder (VAE), harnessing the ability of generative deep learning methods to identify structure in high-dimensional fields through nonlinear dimensionality reduction. The design of the architecture builds on previous work by \cite{spuler_identifying_2024, spuler_learning_2025-1} and \cite{swaroop_learning_2023}. By combining causal inference with deep learning, the DAG-VAE method further enables the generation of data-driven counterfactuals in the climate system. We investigate the ability of this new approach to disentangle the joint influence of the tropical Pacific and Indian Ocean on the short rains over the GHA with the following objectives: 1) Learning interpretable reduced representations of large-scale modes of variability alongside their causal relations; 2) Estimating the spatial causal response patterns of precipitation to individual modes of variability in the causal graph; 3) Generating counterfactuals of seasonal precipitation based on the causally-related reduced representations.\\

Section \ref{data} provides details of the data used to train the model. We then introduce the method, DAG-VAE, in Section \ref{method} and contextualise it in the field of causal representation learning. Section \ref{results} presents the results. We first apply the method to seasonal hindcasts produced by ECMWF’s seasonal prediction system SEAS5 \citep{johnson_seas5_2019}. This allows us to compare our results with the model experiments conducted by \cite{macleod_causal_2021} and to evaluate the robustness of the joint representations on a larger test set than would be available from reanalysis or observational data. We can thereby benchmark the data-driven approach to seasonal attribution presented in this paper to state-of-the-art methods based on model experiments. We then apply the method to reanalysis data - where replacement experiments could not be conducted - and compare the spatial causal effects estimated from the Pacific and Indian Ocean basins to the results found in the seasonal prediction model. In a final step, we generate counterfactual short rain seasons based on combinations of recent extreme drivers. We discuss these results and conclude in Section \ref{discussion}.

\section{Data}\label{data}

We investigate SEAS5 seasonal hindcasts initialized retrospectively from ERA-Interim using the IFS CY43R1 on a TCo319 grid for the time period from 1981 to 2023 \citep{johnson_seas5_2019}. We consider  the short rains season (October - December) for the following variables and regions:
\begin{itemize}
    \item SSTs in the tropical Pacific: (latitudes: $20^\circ S - 20^\circ N$ ; longitudes: $130^\circ E - 75^\circ W$);
    \item SSTs in the Indian Ocean: (latitudes: $10^\circ S - 20^\circ N$ ; longitudes: $40 - 110^\circ E$);
    \item Precipitation over the Greater Horn of Africa: (latitudes: $4.5^\circ S - 11.5^\circ N$ ; longitudes: $29.5 - 50.5^\circ E$).
\end{itemize}

\begin{figure}[!]
\centering
  \noindent\includegraphics[width=38pc,angle=0]{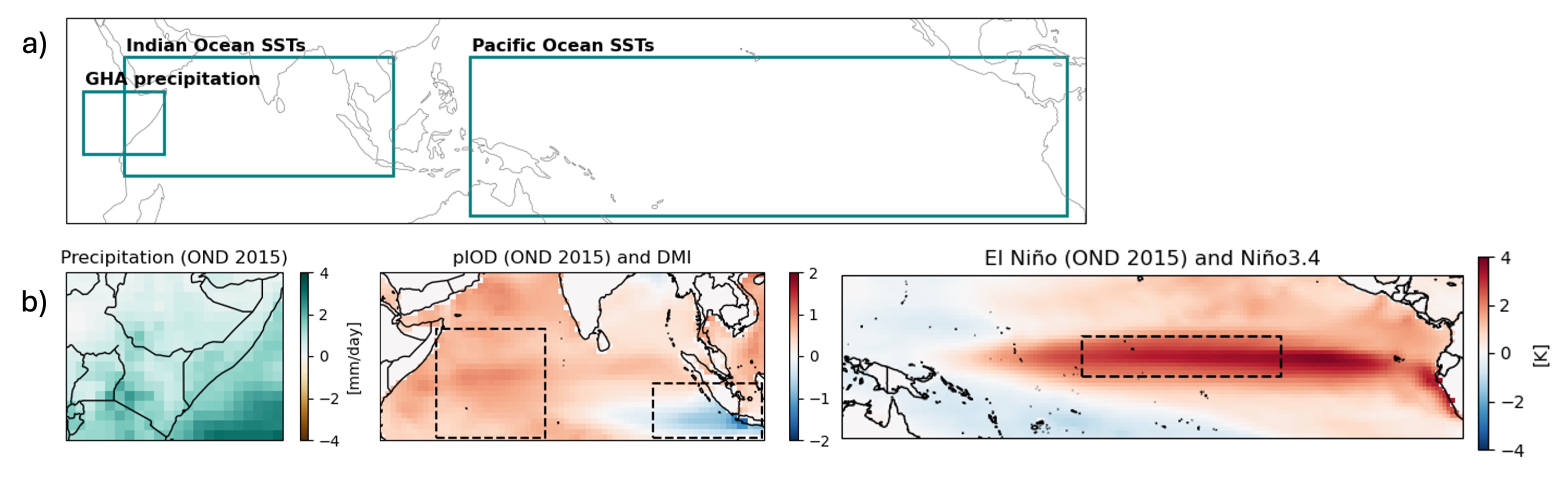}\\
  \caption{a) Illustration of the selected regions and variables used to study Greater Horn of Africa (GHA) precipitation, Indian Ocean (IO) and Pacific Ocean sea surface temperatures (SSTs). (b)Anomalies of precipitation over the GHA, SST patterns associated with a positive Indian Ocean Dipole and an El Niño season, showing October to December averages of the seasonal hindcast initialised in September 2015. The dashed black boxes indicate the regions used to compute the Dipole Mode Index and the Niño3.4 index.}\label{data_figure}
\end{figure}

Both monthly data and seasonal data are analysed in SEAS5, and lead times of 2-6 months are retained for verification months October-December. When analysing seasonal averages, this includes initialization dates from July-September; for individual months, this includes start dates from April-November. Forecast skill of SEAS5 for the short rains is discussed in \citet{tefera_seasonal_2025, nana_assessing_2025}. \\

We investigate the sensitivity of results to using only September start dates and find no major qualitative differences in the results (not shown). Furthermore, the sensitivity to applying the method on monthly vs seasonal data was investigated and no qualitative differences were found in the average causal effect estimated, which is the focus of this paper. We choose to work with seasonal data to more closely match \cite{macleod_causal_2021} to facilitate comparison.\\

The SEAS5 hindcasts are pre-processed in the following way: The mean bias is removed by computing anomalies with respect to the verification month and lead time. Furthermore, we remove the average trend over the area, ensemble, lead time and verification month. We choose to remove the spatially averaged trend to retain pattern effects and their influence on rainfall. The sensitivity of the results to removing the trend, or removing the grid-point-wise trend, is analysed and no major difference in the causal response patterns estimated is found. In the Pacific, the mean estimate of the trend in seasonal hindcasts is found to be lead time dependent (see Appendix A), consistent with the findings of \cite{mayer_tropical_2025}. However, the uncertainty in the estimate of the trend is large compared to the lead time dependence of the mean estimate. We therefore choose to remove the trend averaged over lead times. Furthermore, we standardise the data prior to input to the neural network. Since the grid-point-wise averages are already removed, we only divide by the average standard deviation of the entire field to retain the higher variability in some regions (e.g. the east Indian Ocean dipole region).\\

We fine-tune the architecture on ERA5 reanalysis data \citep{hersbach_era5_2020} from 1981-2023 for the same variables and regions as for SEAS5, whereby ERA5 uses the HadISST2.1.1.0 SST dataset as boundary condition. We regrid ERA5 data to match SEAS5 ($1^\circ$ resolution) and aggregate it from hourly data to 30-day averages, with data every 5 days used for training. The remaining pre-processing mirrors the SEAS5 pre-processing described above.

\section{Method - DAG-VAE}\label{method}

\begin{figure}[!]
\centering
  \noindent\includegraphics[width=38pc,angle=0]{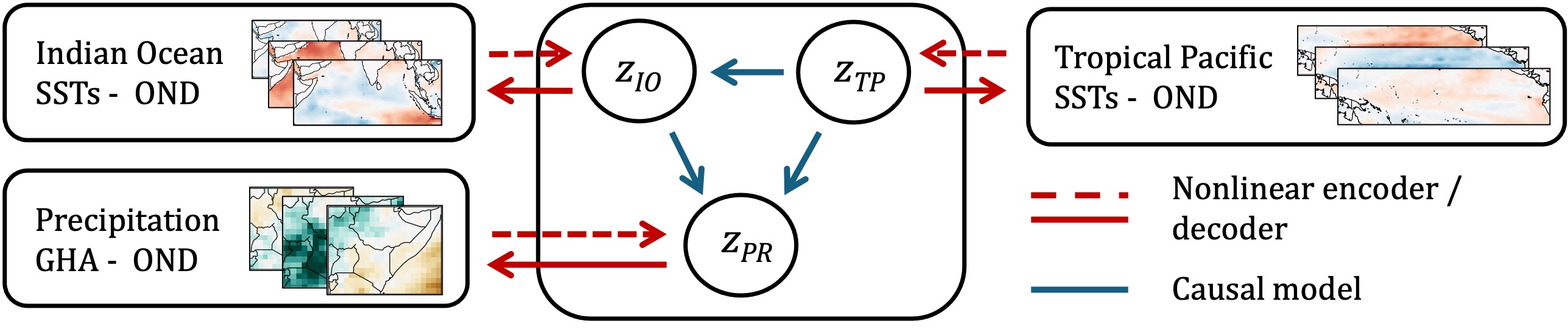}\\
  \caption{Illustration of the DAG-VAE method and architecture. $z_{TP}$, $z_{IO}$ and $z_{PR}$ refer to the low-dimensional representations identified by the model for the three high-dimensional input spaces: tropical Pacific SSTs, Indian Ocean SSTs and GHA precipitation. The red arrows refer to encoder and decoder neural networks that dimensionality-reduce the high-dimensional input space. The blue arrows refer to the directed acyclic graph (DAG) hypothesized in the latent space.}\label{dagvae_illustration}
\end{figure}

We propose the Directed Acyclic Graph Variational Autoencoder (DAG-VAE) method which combines variational autoencoders for nonlinear dimensionality reduction with directed acyclic graphs for causal inference in a single architecture, illustrated in Figure \ref{dagvae_illustration}. Variational autoencoders (VAEs) are a class of generative deep learning models that identify a probabilistic reduced representation of a high-dimensional input space \citep{kingma_auto-encoding_2013}. VAEs enable the generation of new samples from the probability distribution identified in the reduced (also called latent) space, and are trained using Bayesian variational inference, whereby a prior is set on the probability distribution of the latent space. In our application, this prior encodes a DAG between the latent distributions of coupled variational autoencoders. Each node of the DAG represents the latent representation of one of the variables - SSTs over the Indian and Pacific Oceans, and precipitation over the GHA - while each directed edge represents the hypothesized causal influence between variables. The DAG-VAE method builds on earlier work introduced by the authors in \citep{spuler_identifying_2024, spuler_learning_2025-1}. \\

The causality of this model is based on dynamical understanding of the physical processes involved in the teleconnections studied \citep{kretschmer_quantifying_2021}, as well as proofs for identifiability of the model presented by \cite{lachapelle_disentanglement_2022} and \cite{shen_weakly_2022}. Based on existing literature and understanding of dynamical processes discussed in Section \ref{introduction}, we assume a structural causal model in the latent space, shown in Figure \ref{dagvae_illustration}. In particular, we assume that teleconnections from both the tropical Pacific and Indian Ocean are the primary teleconnections influencing the short rains on the timescales and lead times considered and that ENSO is the leading mode of variability on these timescales. Second, we assume that the selected variables, SST and precipitation anomalies, represent the physical processes underlying the teleconnection. \\

The field of causal representation learning was challenged early on by the result that unsupervised learning will not, in general, identify disentangled representations \citep{locatello_challenging_2019} and that several different representations can lead to the same loss function - the 'true' causal relations are said to be non-identifiable in this case \citep{scholkopf_toward_2021}. Identifiability is defined as the uniqueness of the identified model parameters $\theta$ given an observed probability distribution $P_\theta$ ($P_\theta  = P_{\Tilde{\theta}} \Rightarrow \theta= \Tilde{\theta}$). Research effort has since focused on finding strategies to ensure identifiability, often up to transformations such as permutations \citep{khemakhem_variational_2020, lachapelle_disentanglement_2022, brouillard_causal_2024}. \cite{shen_weakly_2022} show that setting a structural causal model as a prior in the latent space ensures disentanglement of the causal factors conditional on the prior assumed. Furthermore, \cite{lachapelle_disentanglement_2022} show that if the target field (precipitation in our case) is affected sufficiently strongly by external observed influences (ENSO and the IOD in our case), then identifiability up to certain transformations can be ensured by enforcing sparsity regularisation. Sufficiently strongly here means that the response to the external factors, called actions in \cite{lachapelle_disentanglement_2022}, covers the full latent subspace that is causally influenced by that action. \\

Building on this research, the DAG-VAE method sets a structural causal model prior based on physical assumptions, and enforces sparsity both in this latent causal model by introducing a LASSO penalty (L1 norm), as well as in the number of latent dimensions in the reduced representations to ensure the disentanglement of the signals of interest in our representations. In this paper, the number of reduced dimensions was set to one for the tropical Pacific and Indian Ocean, and two for monthly precipitation over the GHA, as this was the minimal model needed to capture the dynamical processes analyzed. However, alternative versions with a higher number of latent dimensions (4-10) also produced qualitatively similar precipitation response patterns, as did versions that included independent noise dimensions for precipitation (not shown). \\

The DAG-VAE method jointly fits a nonlinear dimensionality reduction through encoder/decoder neural networks (red arrows in Figure \ref{dagvae_illustration}) and a causal model between the latent representations $z_{TP}$, $z_{IO}$ and $z_{PR}$ (blue arrows in Figure \ref{dagvae_illustration}), which are samples from probability distributions with Gaussian priors. This joint optimisation is conducted through the minimisation of the following modified loss function: 

\begin{align}
    L_{DAG-VAE} (x_{TP},x_{IO},x_{PR}) = &\;
    \mathbb{E}_{q(z_{TP} \mid x_{TP})}  \log p(x_{TP} \mid z_{TP})
    - D_{KL} (q(z_{TP} \mid x_{TP}) \,\|\, p(z_{TP})) \\
    + &\;
    \mathbb{E}_{q(z_{IO} \mid x_{IO})} \log p(x_{IO} \mid z_{IO})
    - D_{KL} (q(z_{IO} \mid x_{IO}) \,\|\, p(z_{IO} \mid z_{TP})) \nonumber \\ 
    + &\;
    \mathbb{E}_{q(z_{PR} \mid x_{PR})} \log p(x_{PR} \mid z_{PR})
    -D_{KL} (q(z_{PR} \mid x_{PR}) \,\|\, p(z_{PR} \mid z_{IO}, z_{TP})) \nonumber
\end{align}

The loss function can be consistently derived from the graphical model shown in Appendix B, alongside a detailed description of the architecture and hyperparameters chosen. The first terms in each row represent the reconstruction loss terms for each of the input fields $x_{TP}$, $x_{IO}$ and $x_{PR}$. The second terms in each row minimize the KL-divergences (denoted $D_{KL}$) - measures of the distance between two probability distributions - between the latent representation based on the input fields and the prediction of the latent representation based on the causal parents. The loss function thereby encodes the dual aim of the model, in identifying meaningful representations of the high-dimensional gridded input data, as well as informative and causally related signals. \\

We choose to model the causal relationships in the latent space using a linear model for a number of reasons, although implementing a nonlinear model would merely require adding an additional layer and activation function to the latent neural network. We assume that nonlinearities in the teleconnections can be captured through the nonlinear dimensionality reduction, enabling us to construct a linear model in the latent space which is sparser (less prone to overfitting) and for which the causal effect estimation is simpler. This is based on the intuition that the reduced representations themselves are affected by less noise (thus, are less prone to overfitting) than the relationships between the full representations. Furthermore, recent results suggest that the combination of nonlinear dimensionality reduction and linear relationships is the most successful when comparing the results of joint representation learning to other approaches \citep{brouillard_causal_2024}.\\

In this version of the architecture, no time-lags are included in the model, although the method can, in principle, be fit for any chosen time-lag. The choice to not include time-lags here was made as the atmospheric response to the SSTs and resulting teleconnection to precipitation over the GHA is assumed to occur at timescales below one month, and is therefore 'instantaneous' when working with seasonal or monthly data. This assumption and choice is in line with current practices in SST replacement experiments such as \cite{macleod_causal_2021}. \\

We compare our results against three baseline methods where appropriate. The first combines linear dimensionality reduction using Principal Component Analysis (PCA, also called EOF analysis in climate science) on the three input fields individually, retaining the same amount of latent dimensions as chosen for the DAG-VAE, with a subsequent linear prediction model.  PCA, commonly referred to as EOF analysis in climate science, projects a higher-dimensional input space into a reduced space spanned by the orthogonal eigenvectors of the covariance matrix of the data \citep{jolliffe_principal_2016}. As discussed in the introduction, PCA is a widely-used method in the study of teleconnections \citep{dommenget_cautionary_2002}. The second and third methods are based on established indices of ENSO and IOD, the Dipole Mode Index for the Indian Ocean and the Niño3.4 index for the Pacific, shown in Figure \ref{data_figure}. In the first version (Indices v1), we use regionally averaged precipitation over the GHA to estimate the teleconnection, as, for example, in \cite{kolstad_lagged_2022}. In the second version (Indices v2), we use these two indices to fit a linear regression to precipitation at each grid point over the GHA.

\section{Results}\label{results}

\subsection{Identifying causal representations in seasonal hindcasts}

\begin{figure}[!]
\centering
  \noindent\includegraphics[width=38pc,angle=0]{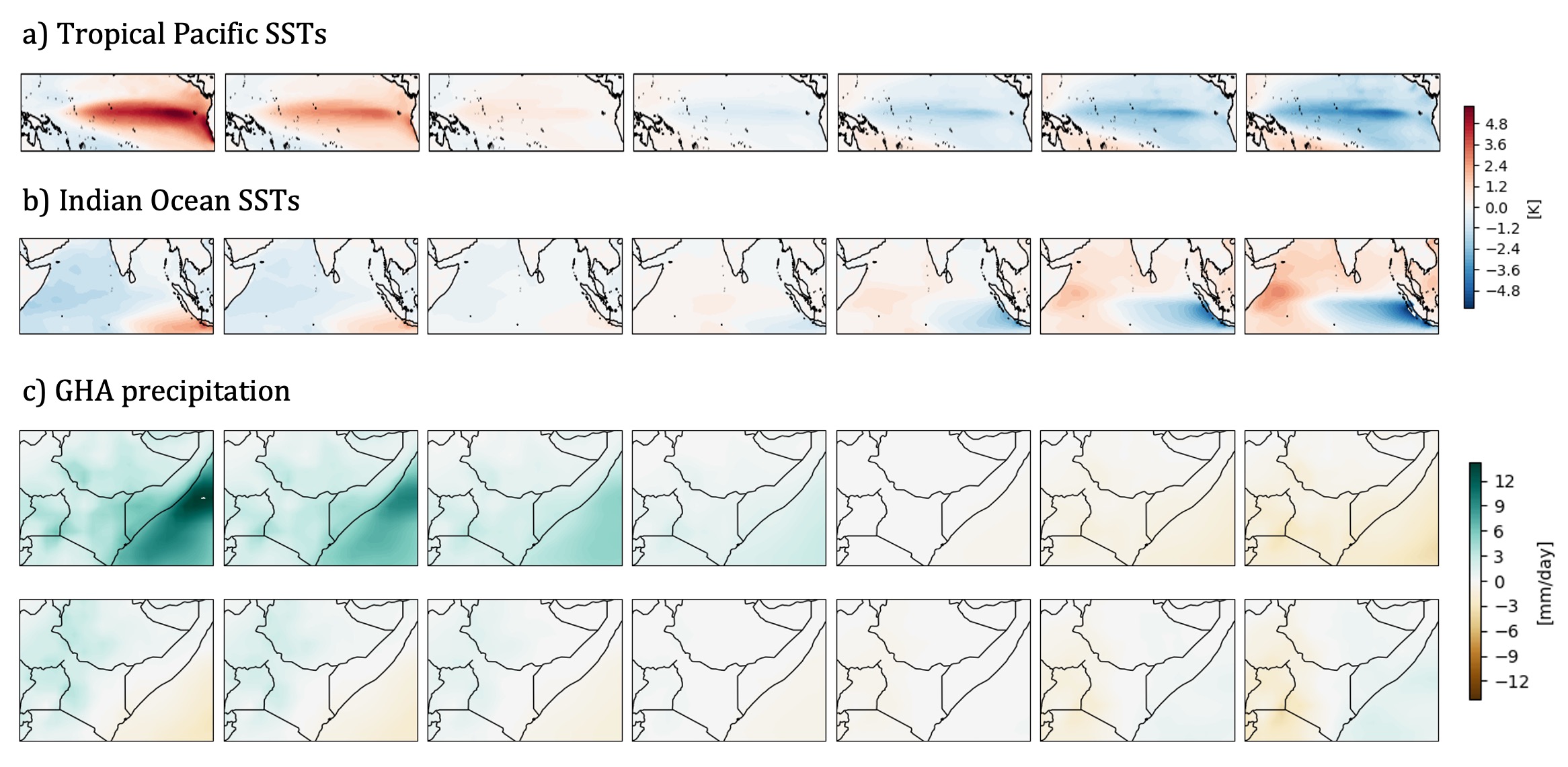}\\
  \caption{Illustration of the learned causal representations in SEAS5, 
  shown through reconstructed samples of the latent dimensions $z_{TP}$, $z_{IO}$ and $z_{PR1,2}$. The samples are created by first selecting equidistant samples from the means of the latent Gaussians, ranging from the minimum to the maximum value encountered in test set. Second, these samples are passed back through the decoder to reconstruct their original dimension.}\label{seas5_representations}
\end{figure}

We first evaluate the ability of the DAG-VAE to identify reduced and informative representations of the modes of variability when trained on seasonal hindcasts. To this end, we first visualize the information encoded in the latent space of the VAE by sampling from the reduced representations of each encoded variable. These samples are then passed through the decoder and thereby reconstructed to their original dimensions. Results are shown in Figure  \ref{seas5_representations}. \\

We find that the DAG-VAE identifies reduced representations $z_{TP}$ and $z_{IO}$ that capture SST patterns associated with the two main modes of variability in the two ocean basins during this season, ENSO and the IOD, seen in the decoded samples of the latent dimension representing the Tropical Pacific SSTs in Figure \ref{seas5_representations}a, and the Indian Ocean SSTs in Figure \ref{seas5_representations}b. This result shows that the DAG-VAE method is able to identify representations that are interpretable and physically meaningful. \\

Furthermore, the nonlinear dimensionality reduction in DAG-VAE is able to capture known asymmetries of both the ENSO and IOD patterns: El Niño anomalies are slightly more intense and reach further east than the La Niña anomalies, which are located more in the central Pacific; positive IOD events are also associated with more intense SST anomalies, in particular the negative anomalies in the east Indian Ocean which also extend spatially further west than the corresponding positive anomalies in the negative IOD phase. The detection of these asymmetric patterns would not be possible with a linear dimensionality reduction method such as PCA. \\

In terms of the reduced representations of precipitation over the GHA, we find that the first dimension, $z_{PR1}$, shown in the first row of Figure \ref{seas5_representations}c, represents an overall wetting/drying over the entire region. Again, there is a slight asymmetry captured by the nonlinear dimensionality reduction: while the drying signal is relatively uniform spatially, the wetting signal is slightly stronger off the coast of the GHA, reflecting a known bias in the seasonal forecasting system. The second dimension, $z_{PR2}$, captures a slightly asymmetric East-West dipole response. \\

\begin{table}[!]
\caption{Evaluation of key characteristics of the latent dimensions. The numbers in brackets refer to the metric evaluated only over the test set. The number after the $\pm$ refers to the standard deviation.}\label{table_seas5}
\begin{center}

\begin{tabularx}{\textwidth}{|X|l|l|l|}
\multicolumn{4}{l}{\textbf{a) Reconstruction loss} - Root Mean Squared Error (RMSE)} \\
\hline
 & Tropical Ocean SSTs & Indian Ocean SSTs & GHA precipitation \\
\hline
 DAG-VAE & $0.32 \pm 0.18$ ($0.33 \pm 0.18$) & $0.51 \pm 0.41$ ($0.52 \pm 0.42$) & $0.31 \pm 0.17$ ($0.32 \pm 0.18$) \\
 \hline
 PCA & $0.36 \pm 0.20$ & $0.64 \pm 0.48$ & $0.36 \pm 0.21$ \\
\hline
\end{tabularx}

\vspace{0.8em} 

\begin{tabularx}{\textwidth}{|l|X|X|X|X|}
\multicolumn{4}{l}{\textbf{b) Predictive performance} - GHA precipitation} \\
\hline
  & DAG-VAE & PCA & Indices v1 & Indices v2 \\
 \hline
Anomaly Correlation Coefficient & 0.68 (0.65) & 0.51 (0.50) & 0.39 (0.38) & 0.50 (0.48)\\
 \hline
 $R^2$ total precipitation & 0.58 (0.55) & 0.43 (0.43) & 0.22 (0.20) & 0.41 (0.39) \\
 \hline
\end{tabularx}

\vspace{0.8em} 

\begin{tabularx}{\textwidth}{|l|X|X|X|X|}
\multicolumn{5}{l}{\textbf{c) Statistical robustness} - Mean correlation coefficient (MCC) computed on the test set} \\
\hline
 & $z_{TP}$ & $z_{IO}$ & $z_{PR1}$ & $z_{PR2}$ \\
\hline
 DAG-VAE & $0.97 \pm 0.03$ & $0.90 \pm 0.05$ & $0.93 \pm 0.08$ & $0.90 \pm 0.06$ \\
 \hline
\end{tabularx}

\end{center}
\end{table}

Next, we evaluate the performance of the DAG-VAE in comparison to the three baselines introduced in Section \ref{method}: Principal component analysis combined with regression (PCA), indices with a regression on regional average precipitation (Indices v1), and indices with grid-point-wise regression (Indices v2). Results are shown in Table \ref{table_seas5}. We first assess how well the reduced representations capture variability in the original input space, quantified using the root mean squared error (RMSE) of the reconstructed data with respect to the original input data (Table \ref{table_seas5}a). Since the indices cannot reconstruct the original space as such, the DAG-VAE is only compared to PCA here. Here, the DAG-VAE outperforms PCA, i.e. the reconstruction loss is lower. This result is in line with the visualisation of the latent dimensions shown in Figure \ref{seas5_representations}, which showed that the nonlinear dimensionality reduction largely captures the main modes of variability but is able to better represent the spatial asymmetries. Both methods show a large variance in the reconstruction loss, quantified by the standard deviation shown in Table \ref{table_seas5}. This is due to the imposed constraint of one/two dimensional representation, which will not be able to reconstruct patterns of SST anomalies not directly linked to the main mode of variability.\\

We furthermore assess the ability of the reduced dimensions of the tropical Pacific and Indian Ocean to predict precipitation over the GHA, evaluated using the Anomaly Correlation Coefficient, as well as the $R^2$ of total precipitation over the region. $R^2$ is defined as the ratio of the residual sum of squares to the total sum of squares subtracted from 1. With anomaly correlations around r = 0.68, DAG-VAE outperforms both PCA (r = 0.51) and indices with regionally averaged precipitation (r = 0.39) or grid-point-wise regression (r=0.50) in terms of the ability of the reduced representations to predict precipitation, on both the full and test data. This improved performance can be attributed to the joint learning of reduced representations and causal inference, highlighting the potential of both the approach in general and this method specifically to find predictable components of seasonal teleconnections. \\

The statistical robustness of the DAG-VAE representations is assessed through the mean correlation coefficient of the latent dimensions identified when training the neural network multiple times \citep{brouillard_causal_2024} which is high, indicating that the mean of the latent dimensions is statistically robust and that the identifiability argued based on \cite{shen_weakly_2022} and \cite{lachapelle_disentanglement_2022} holds empirically. \\

Overall, the DAG-VAE method identifies reduced representations that are both dynamically interpretable and improve upon existing linear approaches in terms of reconstructing the input space and predicting the target variable, precipitation.

\subsection{Estimating causal response patterns in seasonal hindcasts}\label{section_seas5_causal}

Based on the evaluation conducted in the previous section, we now use the DAG-VAE model trained on seasonal hindcasts to disentangle the direct causal effects of variability in each of the two ocean basins on the GHA short rains. To this end, we simulate intervention experiments in the latent space of the DAG-VAE, setting one of the ocean basins to climatology and evaluating the effect of variability in the other. \\

This approach of setting one ocean basin to climatological conditions resembles the experimental set-up in \cite{macleod_causal_2021}, who study the effect of both realistic and box heating experiments in each of the basins using the ECMWF Integrated Forecasting System (IFS), cycle CY41R1 at T255, against which we benchmark our results. \cite{macleod_causal_2021} find wet/dry conditions over the entire GHA in response to isolated SST replacement in the Indian Ocean, and a precipitation dipole in response to isolated SST replacement in the Indian Ocean with wet anomalies off the coast and dry anomalies inland in response to El Niño (opposite for La Niña).  Given that we are using the same forecasting model, albeit of a slightly later version, we would expect to find similar results to the SST replacement experiments, so long as our assumptions about the physical mechanisms involved are correct. \\

\begin{figure}[!]
\centering
  \noindent\includegraphics[width=38pc,angle=0]{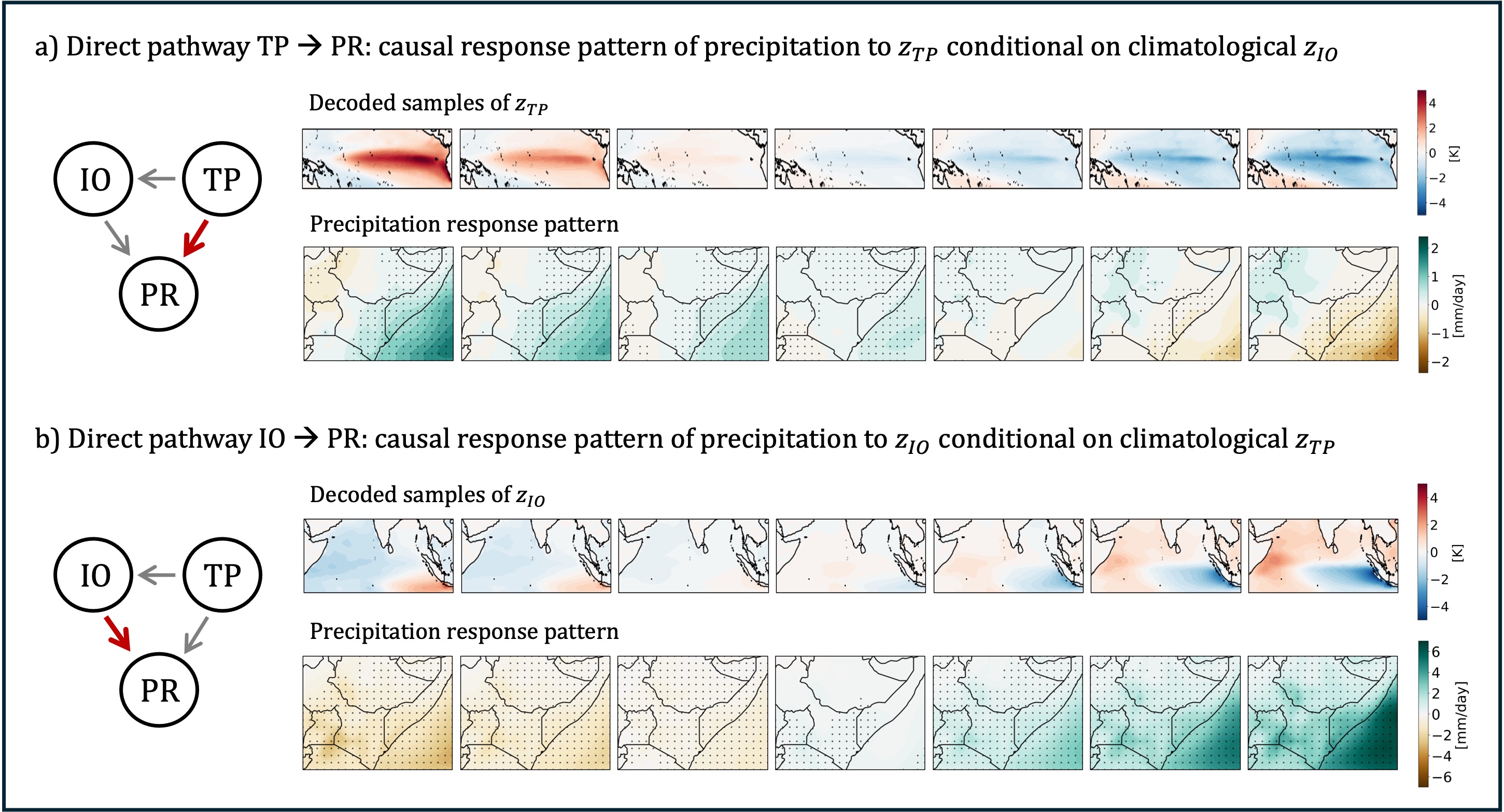}\\
  \caption{Intervention experiments conducted using DAG-VAE trained on SEAS5. Stippling indicates anomalies whose sign is consistent across 25 training runs, sampling the uncertainty introduced through the initial weights. a) The latent variable representing SSTs in the Indian Ocean, $z_{IO}$, is set to climatological (neutral) conditions, and samples are drawn from the latent variable representing SSTs in the Pacific, $z_{TP}$, shown in the top row after decoding. The effect of this intervention on precipitation over the GHA is estimated using the causal model fit in the latent space and shown in the bottom row, aligned to the corresponding sample s of $z_{TP}$. b) Here, $z_{TP}$  is set to neutral conditions and samples are drawn from $z_{IO}$ (top row), with the corresponding effect on precipitation over the GHA shown in the bottom row.}\label{seas5_causal}
\end{figure}

To estimate the direct effect of the Indian Ocean on precipitation over the GHA, we set the Pacific to climatological conditions defined as the absence of anomalies, and estimate precipitation response patterns based on sampling conditions in the Indian Ocean. These response patterns are estimated using the causal graph fit in the latent space of the model, and represent the spatial causal effect estimated by the DAG-VAE. We proceed equivalently to quantify the direct effect of the Pacific by fixing the Indian Ocean to climatological conditions and estimate precipitation response patterns to variations in the Pacific. Results for both experiments are shown in Figure \ref{seas5_causal}.\\

We find that the DAG-VAE is able to recover the results of the SST replacement experiments: the direct effect of a positive IOD event is a wet anomaly over the entire region, whereas the effect of a negative IOD event is a region-wide dry anomaly (Fig \ref{seas5_causal}b). The direct effect of ENSO is weaker (note the different colorbar range), and the precipitation response is an east-west dipole with dry anomalies inland and wetting off the coast in response to El Niño, and vice-versa in response to La Niña (Fig. \ref{seas5_causal}a). Further evaluation of the mechanism underlying these two distinct precipitation response patterns conducted by \cite{macleod_causal_2021} finds that without the additional IOD anomaly, zonal wind anomalies induced by El Niño extend further inland, inhibiting deep convection associated with the pIOD, but inducing off-coast advection-driven positive rainfall anomalies. \\

Overall, we show that the DAG-VAE method is able to attribute different precipitation response patterns to variability from different ocean basins, recovering results from SST replacement experiments conducted by \cite{macleod_causal_2021} which represent a widely-used approach to studying the causal effects of teleconnections. The ability of the DAG-VAE trained on seasonal hindcasts to both outperform existing linear methods in terms of reconstruction and prediction, as well as recover results of SST replacement experiments, further shows the benefit of jointly identifying physically interpretable reduced representations and their teleconnections in a data-driven causal framework. Disentangling these two distinct precipitation response patterns, and in particular recovering the dipole response to ENSO, would not have been possible using indices, such as the Niño3.4 index, Dipole Mode Index and regionally averaged precipitation over the GHA used, for example, in \cite{kolstad_lagged_2022}, as the average of the precipitation dipole response pattern could not have been distinguished from no anomaly. In approaches such as PCA combined with lasso regression or grid-point wise regression, the identified patterns would have been linear, i.e. could not have captured asymmetry in the opposite responses to positive/negative ENSO or IOD events, and would have explained less spatial variability.

\subsection{Causal representations and response patterns learned in reanalysis data}

So far, we have trained and evaluated the DAG-VAE on the ensemble of seasonal hindcasts. This allowed us to sample a wider range of plausible ENSO and IOD events and their respective response patterns, and compare our results directly to the SST replacement experiments. However, besides being computationally less expensive, a key benefit of the data-driven approach to studying causal response patterns presented here is that it can also be applied to observations or reanalysis data. \\

We now train the model on ERA5 reanalysis data and compare both the reduced representations and the causal response patterns identified to those found in SEAS5. We implement this training as a fine-tuning of the model weights trained on SEAS5 data as specified in Section \ref{method}. This is based on the assumption that although the seasonal hindcasts have known biases in both modes of variability and their teleconnections, the reanalysis-based representations can be found in the neighbourhood of the seasonal model-based representations. As for SEAS5, the model is trained 25 times to investigate uncertainty across initial conditions. \\

\begin{figure}[!]
\centering
  \noindent\includegraphics[width=38pc,angle=0]{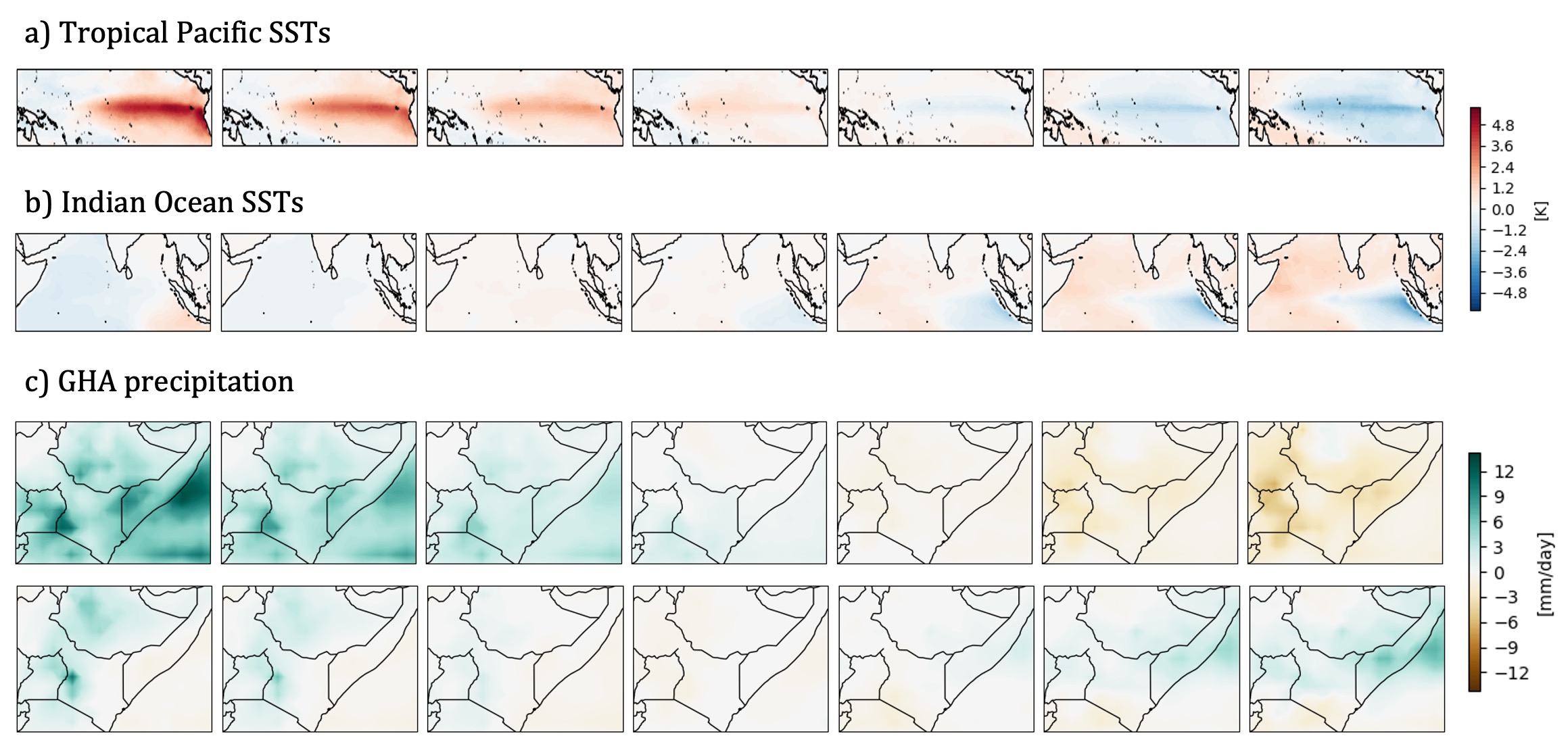}\\
  \caption{Same as Figure \ref{seas5_representations}, but for ERA5.}\label{era5_representations}
\end{figure}

Figure \ref{era5_representations} shows the causal representations identified in reanalysis data. While qualitatively consistent, we find slight differences in the representations identified in ERA5 compared to those identified in SEAS5, shown in Figure \ref{seas5_representations}, highlighting known biases in SEAS5. Both ENSO and the IOD exhibit a higher positive skewness in reanalysis data compared to seasonal hindcasts. Furthermore, the positive SST anomalies associated with El Niño, and the negative SST anomalies associated with a positive Indian Ocean Dipole extend less far east. The reduced representations of precipitation show more spatial disaggregation, and, in the case of the first dimension, less strong wet anomalies over the ocean but stronger anomalies over the land. Furthermore, the second latent precipitation dimension no longer shows a symmetric East-West dipole as it did for SEAS5.\\

Table \ref{table_era5} again compares the DAG-VAE to the three baselines: PCA-regression (PCA), index-based regression to regionally averaged precipitation (Indices v1) and grid-cell wise index-based regression (Indices v2). As in SEAS5, the DAG-VAE outperforms both PCA and index regressions in terms of reconstruction loss and predictive performance. In all three baseline methods, the performance of the prediction drops sharply when evaluated over the chosen testing period (1981-1990), attaining negative $R^2$ values. While the performance of the DAG-VAE method also drops, it retains a higher performance in both the ACC and $R^2$, indicating a higher statistical robustness of the method in this application.\\

\begin{table}[!]
\caption{Same as Table \ref{table_seas5}, but for ERA5.}\label{table_era5}
\begin{center}

\begin{tabularx}{\textwidth}{|X|l|l|l|}
\multicolumn{4}{l}{\textbf{1) Reconstruction loss} - Root Mean Squared Error (RMSE)} \\
\hline
 & Tropical Ocean SSTs & Indian Ocean SSTs & GHA precipitation \\
\hline
 DAG-VAE & $0.25 \pm 0.16 \; (0.23 \pm 0.16)$  & $0.47 \pm 0.38 \; (0.45 \pm 0.37)$ & $0.29 \pm 0.22 \; (0.31 \pm 0.24)$ \\
 \hline
 PCA & $0.32 \pm 0.21$ & $0.62 \pm 0.56 $ & $0.40 \pm 0.30$  \\
\hline
\end{tabularx}

\vspace{0.8em} 

\begin{tabularx}{\textwidth}{|l|X|X|X|X|}
\multicolumn{4}{l}{\textbf{2) Predictive performance} - GHA precipitation} \\
\hline
  & DAG-VAE & PCA & Indices v1 & Indices v2 \\
 \hline
Anomaly Correlation Coefficient & 0.53 (0.27) & 0.48 (0.17) & 0.44 (0.2)  & 0.49 (0.21) \\
 \hline
 $R^2$ total precipitation & 0.58 (0.47) & 0.43 (-0.63) & 0.48 (-0.26) &  0.48 (-0.25) \\
 \hline
\end{tabularx}

\vspace{0.8em} 

\begin{tabularx}{\textwidth}{|l|X|X|X|X|}
\multicolumn{5}{l}{\textbf{3) Statistical robustness} - Mean correlation coefficient (MCC) across model runs on test set } \\
\hline
 & $z_{TP}$ & $z_{IO}$ & $z_{PR1}$ & $z_{PR2}$ \\
\hline
 DAG-VAE & $0.98 \pm 0.02$ & $0.87 \pm 0.07$ & $0.98 \pm 0.01$ & $0.81 \pm 0.13$ \\
 \hline
\end{tabularx}

\end{center}
\end{table}

\begin{figure}[!]
\centering
  \noindent\includegraphics[width=38pc,angle=0]{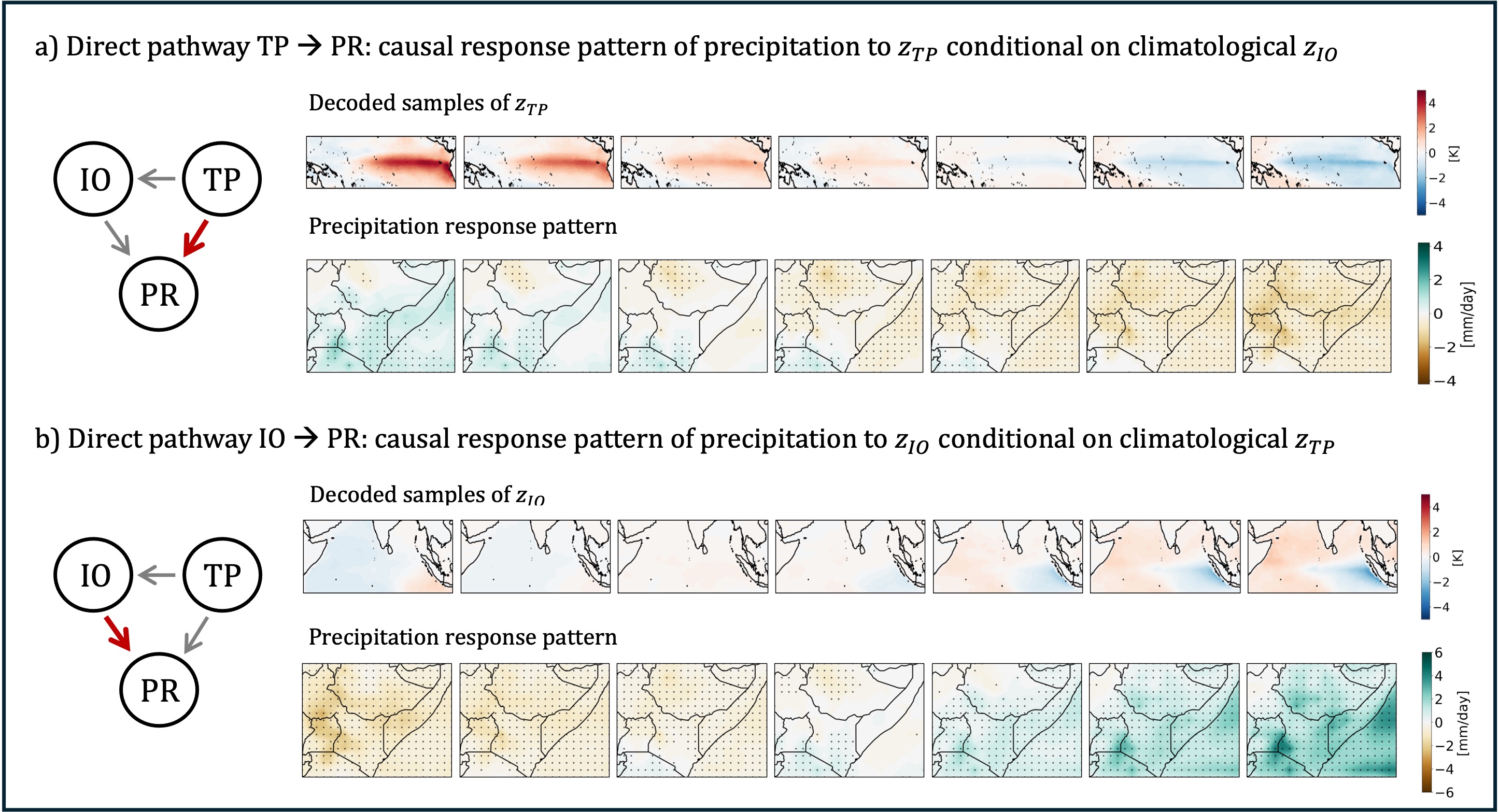}\\
  \caption{Same as Figure \ref{seas5_causal} but for ERA5: intervention experiments conducted using DAG-VAE trained on ERA5. Stippling indicates anomalies whose sign is consistent across 25 training runs.}\label{era5_causal}
\end{figure}

Next, we estimate the direct causal response patterns in precipitation to teleconnections from the two ocean basins, using the approach presented in Section \ref{results}\ref{section_seas5_causal}. Results are shown in Figure \ref{era5_causal}, in the same format as the results for SEAS5 presented in Figure \ref{seas5_causal}. We find that the direct response of precipitation to the IOD is qualitatively similar in ERA5 compared to SEAS5, a region-wide wetting/drying response. However, the wet response is less intense than in SEAS5 over the ocean and the spatial patterns are slightly different and in some cases nonlinear: the regions with the strongest wet response are located further South and over the ocean, compared to the region with the strongest dry response which is located more towards the East and North, over Uganda and South Sudan. The North-East corner of the selected region over South Sudan furthermore does not seem to exhibit a detectable wet response to El Niño, but exhibits a dry response to La Niña. \\

The direct response to variability in the tropical Pacific, however, differs in ERA5 compared to SEAS5. In SEAS5, the direct response to ENSO is a precipitation dipole, consistent with the results of \cite{macleod_causal_2021}. In ERA5, on the other hand, we do not recover this dipole. Rather, we find a region-wide dry anomaly in response to La Niña with a similar spatial pattern but weaker amplitude (note different axes) as the response to a negative IOD event. The direct precipitation response to El Niño is a slightly weaker wet anomaly located inland over Kenya, Uganda and North Tanzania. This pattern was not shown to be sensitive to different train-test splits of the data, training over the entire period considered in this paper (1981-2023), or training on reanalysis data extending back to 1940, although the region with a consistent response across model runs (indicated through stippling in Figure \ref{era5_causal}) shifts slightly (not shown).  \\

Overall, this section shows that the DAG-VAE method introduced here can be applied to reanalysis or observational data and recover consistent representations and response patterns across model runs. Consistent with SEAS5, the method outperforms linear baseline methods in terms of both reconstruction and prediction, and recovers a spatial precipitation response that is statistically more robust when evaluated over a period that the model was not trained on. The difference in the response patterns between SEAS5 and ERA5, discussed further in Section \ref{discussion}, highlights the benefit of data-driven approaches for analysing the causal effects of joint teleconnections and can be used to diagnose and correct biases in seasonal forecasts as well as climate projections.

\subsection{Data-driven counterfactuals of seasonal precipitation based on reanalysis data}

\begin{figure}[!]
\centering
  \noindent\includegraphics[width=36pc,angle=0]{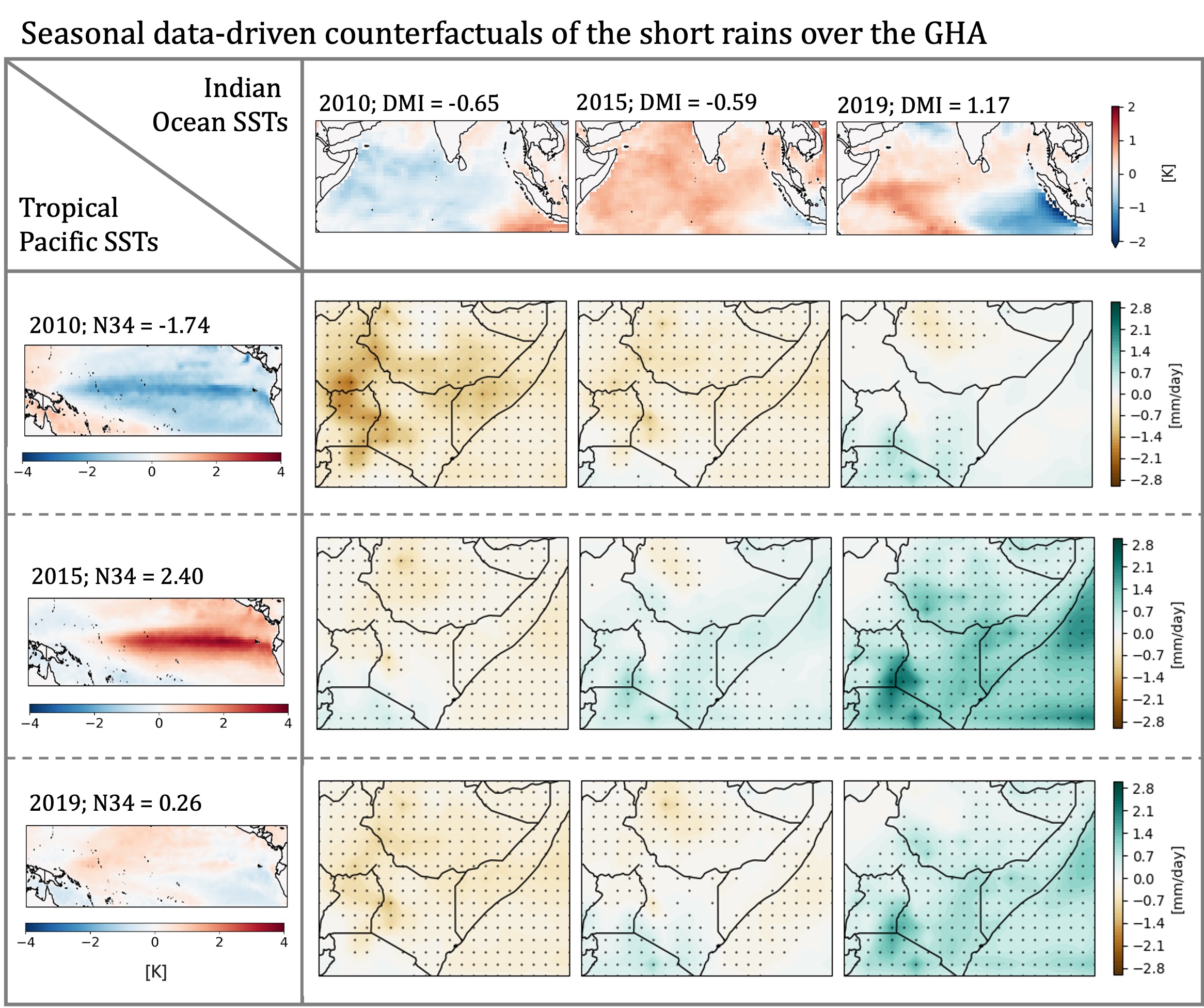}\\
  \caption{Factual and counterfactual anomalies for three recent short rain seasons in the years 2010, 2025, and 2019 based DAG-VAE trained on ERA5. Precipitation anomalies on the diagonal indicate predicted anomalies given the actually observed conditions. Additive counterfactuals are created by encoding the respective ENSO and IOD anomalies predicted in September into the latent space of the DAG-VAE, and predicting the precipitation response using the causal graph. Stippling indicates anomalies whose sign is consistent across 25 training runs, sampling the uncertainty introduced through the initial weights.}\label{counterfactual}
\end{figure}

Having attributed precipitation response patterns over the GHA to variability in the two ocean basins individually, we now use our causal model to generate seasonal counterfactuals. A counterfactual provides the answer to a `what would have happened if' question, such as what would have happened if the strong El Niño of 2015 had co-occurred with the strong positive IOD of 2019. Here, we study the counterfactual effects of recent extreme ENSO and IOD events, focusing on the short rains in 2010 (La Niña, negative IOD), 2015 (strong El Niño, moderately positive IOD) and 2019 (neutral ENSO, strong positive IOD). Figure \ref{counterfactual} shows the precipitation anomalies resulting from combining the anomalies from the tropical Pacific and Indian Ocean basins from the selected years and estimating their joint effect on precipitation using the causal graph in the latent space of the DAG-VAE trained on ERA5. \\

Generating counterfactuals requires specifying the noise terms of the model to the event - in this case, the season - in question. In our context, the noise terms are all other processes driving precipitation over the GHA, including subseasonal drivers such as the Madden-Julian Oscillation. Based on the linearity between the predictive relationships in the latent space, counterfactual states in the two ocean basins are assumed to be additive to these other processes. Figure \ref{counterfactual} therefore shows the additive counterfactual effect introduced through the interaction of different seasonal drivers. Here, we do not quantify the probability of these different states of the Indian and tropical Pacific Oceans co-occurring. \\

In the three example seasons investigated here, we find that the dry anomaly associated with the 2010 La Niña would have still been a dry anomaly if it had co-occurred with the moderate positive IOD of 2015, whereas the response would have been spatially disaggregate if it had co-occurred with the strong positive IOD of 2019. If the 2019 IOD, on the other hand, had co-occurred with the strong El Niño event of 2015, the resulting precipitation anomalies would have been even stronger than those actually observed in 2019 which led to flooding and already ranked among the wettest historically observed short rain seasons. In turn, the negative IOD event of 2010 would have caused wide-spread dry anomalies irrespective of the co-occurring ENSO event, albeit of different magnitudes. \\

In summary, the counterfactual analysis presented here permits the generation of plausible but unseen extreme short rain seasons consistent with large-scale drivers, with possible applications for forecast-based action and scenario planning.

\section{Discussion and Conclusion}\label{discussion}

This paper presents the DAG-VAE method for jointly dimensionality reducing high-dimensional processes in the climate system and disentangling their causal relations. Combining causal inference with deep learning, the method identifies causal relations in a reduced space using variational inference, building on targeted dimensionality reduction approaches developed in \cite{spuler_identifying_2024, spuler_learning_2025-1} and recent progress in the field of causal representation learning. We apply the method to address the question of whether a direct influence from the tropical Pacific on the short rains over the Greater Horn of Africa (GHA) exists, or whether the teleconnection is mediated entirely by the response of the Indian Ocean. \\

The architecture is first trained on seasonal hindcasts and subsequently fine-tuned on ERA5 reanalysis. In both cases, the DAG-VAE method outperforms three linear baseline methods in terms of representing variability of the high-dimensional input fields, as well as in the out-of-sample predictive skill of precipitation given its causal parents (Tables \ref{table_seas5} and \ref{table_era5}). Visualising the respective latent spaces, we find that the DAG-VAE method recovers known modes of variability in the tropical Pacific and Indian Ocean, the El Niño Southern Oscillation (ENSO) and the Indian Ocean Dipole (IOD) (Figures \ref{seas5_representations} and \ref{era5_representations}). Due to the nonlinearity of the dimensionality reduction, the DAG-VAE method is able to capture the spatial assymmetry between positive and negative ENSO and IOD patterns in a single latent dimension. We find that compared to ERA5, SEAS5 appears to underestimate the positive skewness of ENSO and the IOD, and overestimate the amplitude of the IOD, consistent with previous findings \citep{timmermann_ninosouthern_2018, gler_systematic_2025}. \\

Next, we attribute the direct causal response pattern of precipitation to teleconnections from the two ocean basins using the reduced-order model identified by the DAG-VAE. In a data-driven model experiment, we constrain one of the ocean basins to climatological conditions and simulate the precipitation response to variability in the other basin. Trained on SEAS5, we find that DAG-VAE is able to recover the results found in SST replacement experiments conducted by \cite{macleod_causal_2021}: a region-wide wetting/drying response to direct influence from the IOD, and a precipitation dipole response of weaker amplitude in direct response to ENSO (Figure \ref{seas5_causal}). The ability of the proposed approach to identify a response pattern consistent with SST replacement experiments provides evidence for the suitability of the approach, as well as the causal graph hypothesized in this application. \\

However, evaluating the direct effects from the two ocean basins in ERA5 reanalysis data, we do not find evidence for a dipole response pattern to ENSO. Rather, the direct response to La Niña is a region-wide dry response of weaker amplitude than the direct response to a negative Indian Ocean Dipole, and the response to El Niño emerges as a weak wet response primarily over the South-West of the region. This difference between SEAS5 and ERA5 can be due to model biases in the seasonal hindcasts, as well as sample size issues in ERA5. Considering model biases, \cite{macleod_causal_2021} discuss a too-strong response of the IOD to variability in the tropical Pacific, as well as mean-state biases in precipitation and the zonal winds over the Indian Ocean, both of which could influence the precipitation signal attributed to the Pacific Ocean \citep{mayer_assessment_2024}. Investigating potential sources of this discrepancy further is beyond the scope of this paper and will be left to future work. Considering sample size issues, the consistency of the identified pattern when training over different time periods provides evidence for the statistical robustness of this result. However, other sensitivities could remain, introduced for example through the choice of dataset. In particular, recent publications find substantial differences across observational and reanalysis products over the Indian Ocean even after 1990 \citep{wang_indian_2024, jain_towards_2025}, as well as precipitation biases in ERA5 reanalysis data over the region \citep{steinkopf_verification_2022}. \\

Finally, we condition on extreme ENSO and IOD states observed in the recent past to simulate counterfactual short rain seasons (Figure \ref{counterfactual}). We find, for instance, that the short rain season of 2019 which led to wide-spread flooding could have been worse, if an El Niño event of the magnitude and spatial pattern of 2015 had occurred in this year. Overall, the counterfactual analysis presented here exemplifies an approach to generating extreme short rain seasons consistent with large-scale drivers that could be used for scenario planning for unseen events in the humanitarian sector and government.\\

In terms of the dynamical mechanisms considered, we focus on the average causal effects at seasonal timescales in this paper. However, it is known that these seasonal teleconnections are influenced by variability on other timescales that could be explored in future work: in particular, they are modulated by decadal variability such as the Pacific Decadal Oscillation and anthropogenic climate change, and themselves modulate subseasonal teleconnections such as the Madden-Julian Oscillation which is known to influence precipitation over the GHA \citep{clark_interdecadal_2003, kebacho_large-scale_2021, nana_diverse_2025, wainwright_extreme_2021}. The present analysis must therefore be interpreted as estimating causal effects conditional on the timescale (seasonal) and season (October to December) studied. Furthermore, the counterfactual analysis conducted here is conditional on the assumption that the effect of ENSO and the IOD is additive to other drivers of extreme precipitation over the GHA such as the MJO. In addition, the Indian Ocean Dipole is known to peak between September and November, implying that the Indian Ocean Basin Mode, which peaks later in the year, might influence the results we find for October to December \citep{cai_pantropical_2019}. We conducted a sensitivity analysis excluding the month of December from our data and found that the results do not change qualitatively (not shown).\\

The method as it is currently implemented is suitable for studying causal processes at predefined time lags \citep{kretschmer_quantifying_2021}. In future work, the method could be extended to study interactions between multiple time-lags that are not hypothesized apriori \citep{girin_dynamical_2021, boussard_towards_2023}. Furthermore, potential synergies with existing linear approaches in climate science that aim to infer governing equations for high-dimensional systems could be explored. For example, linear inverse models and the associated principal oscillation patterns aim to identify deterministic linear dynamics subject to stochastic forcing within the often linearly dimensionality-reduced space of principal components \citep{kido_understanding_2023, hasselmann_pips_1988, kwasniok_linear_2022, albers_priori_2019}. Furthermore, a range of dynamical systems-based approaches has been used to study reduced representations of teleconnections in the climate system \citep{ghil_review_2023}. In addition, deep learning methods have, in recent years, led to significant advances in the field of equation discovery \citep{brunton_discovering_2016, brunton_promising_2024}, the application of which in climate science has been discussed by \cite{huntingford_potential_2025}. \\

The latent representations identified by this architecture are inherently interpretable and do not require application of explainable AI (XAI) methods such as Shapley values which are known to show architecture-dependent performance differences \citep{bommer_finding_2024}. Here, the choice is made deliberately to instead constrain the latent space and jointly optimise its representativeness and prediction skill to discover causal connections in the latent space, rather than optimise predictive performance for which other deep learning approaches would be more suitable. \\

Overall, we present a data-driven approach to study teleconnections based on observational data that can be used complementary to model experiments. In this paper, we demonstrate the ability of the method to estimate spatial causal response patterns and generate data-driven counterfactuals at seasonal timescales. The findings of this paper could find application in the development of early-warning based action in response to ENSO or IOD forecasts over the Greater Horn of Africa, as well as in the planning for scenarios or storylines of plausible worst-case events \citep{kelder_how_2025}. Furthermore, the spatial response patterns identified in reanalysis data could be used to further investigate and correct model biases in seasonal forecasts \citep{spuler_ibicus_2024}. More broadly, the issue of investigating teleconnections between high-dimensional spatiotemporal processes and the pitfalls of dimensionality-reducing processes prior to studying their teleconnections, has been noted in a wide range of publications \citep{capotondi_understanding_2015, mindlin_plausible_2023, breul_revisiting_2023}, highlighting the potential of the DAG-VAE method to application in other domains of climate dynamics research. \\


\textbf{Acknowledgements:} This research has been supported by the University of Reading (Advancing the Frontiers of Earth System Prediction (AFESP) Doctoral Training Programme) as well as the European Commission, Horizon Framework Programme (XAIDA (Extreme Events: Artificial Intelligence for Detection and Attribution), grant agreement no. 101003469 and EXPECT (Towards an Integrated Capability to Explain and Predict Regional Climate Changes), grant agreement no. 101137656). The authors thank Jakob Wessel for useful discussions and feedback.

%
%

\textbf{Data availability statement:} ERA5 reanalysis data \citep{hersbach_era5_2020} and SEAS5 seasonal hindcast data  \citep{johnson_seas5_2019} were downloaded from the Copernicus Climate Change Service (C3S) (2025). The results contain modified Copernicus Climate Change Service information 2020. Neither the European Commission nor ECMWF is responsible for any use that may be made of the Copernicus information or data it contains. The data underlying the findings is published on Zenodo \href{doi:10.5281/zenodo.17779596}{doi:10.5281/zenodo.17779596}.

\clearpage

\section{Appendix A: Trend exploration in seasonal hindcasts}\label{appendixA}

\begin{figure}[!htp]
\centering
  \noindent\includegraphics[width=38pc,angle=0]{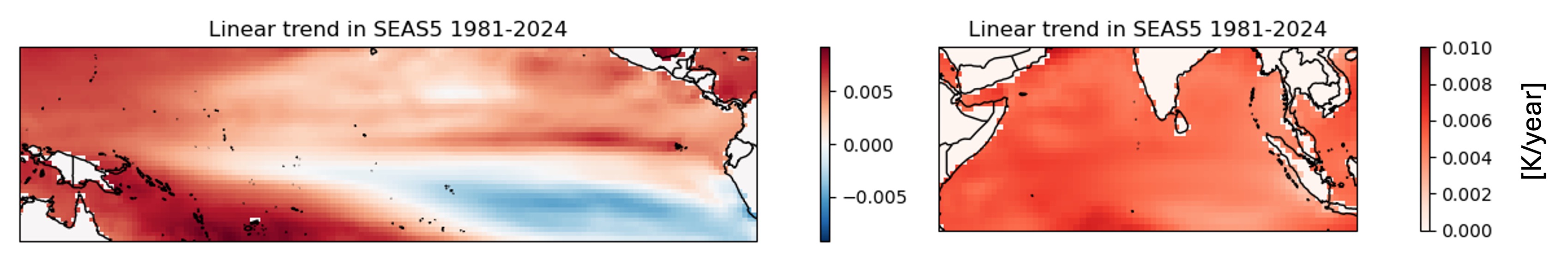}\\
  \caption{Gridpoint wise trend in SEAS5 averaged across leadtimes.}\label{trendplot2}
\end{figure}

Both seasonal and reanalysis data are detrended prior to studying the seasonal teleconnections. We subtract the spatially averaged trend to retain the spatial structure of the trend as it influences the teleconnection studied. Training the model with either non-detrended data or grid-point-wise detrended data does not change the results qualitatively, meaning that the precipitation response patterns shown in Figures \ref{seas5_causal} and \ref{era5_causal} remain similar, and the method still outperforms the baselines investigated (Tables \ref{table_seas5} and \ref{table_era5}). The grid-point-wise linear trend in SEAS5 is shown in Figure \ref{trendplot2}. \\

\begin{figure}[!htp]
\centering
  \noindent\includegraphics[width=38pc,angle=0]{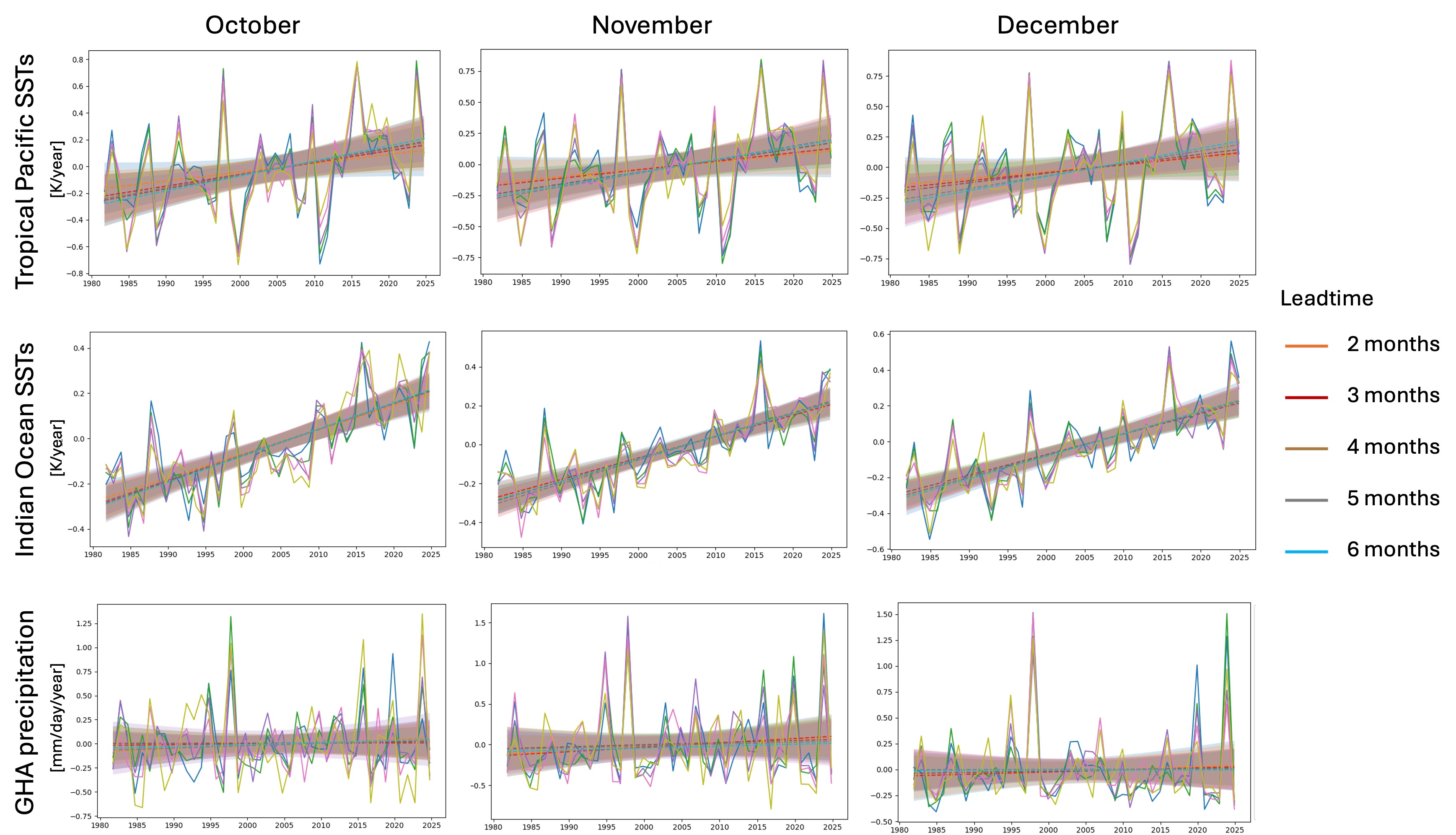}\\
  \caption{Leadtime dependence of area average trend in SEAS5.}\label{trendplot1}
\end{figure}

We also investigate the dependence of the area-averaged trend on the leadtime of the seasonal hindcast, based on recent findings that the El Niño-like trend pattern emerges at later leadtimes in seasonal hindcasts \citep{mayer_tropical_2025}. Results are shown in Figure \ref{trendplot1}. We find that trend of the ensemble mean indeed varies with leadtime over the tropical Pacific (dashed lines). However, bootstrapping this estimate across ensemble members, we find that the uncertainty in the trend estimate is large compared to the difference in the means. We therefore decide to subtract the mean trend across leadtimes, rather than a separate trend for each leadtime.

\section{Appendix B: Experiments, loss function and architecture details}\label{appendixA}

A range of implementations of the architecture were developed and tested. We used three separate encoder/decoder structures to identify physically interpretable latent spaces that would enable separate interventions in each of them. In experiments conducted with a single latent space for all three variables, we found that the modes in the Indian and tropical Pacific Ocean collapsed, and that the reconstruction loss for precipitation compared unfavorably to the model version with separate latent spaces, presumably due to the different properties of its probability distribution. \\

We also tested architectures with higher-dimensional latent spaces (4-10 dimensions per variable) and found qualitatively similar response patterns, however without being able to guarantee identifiability based on \cite{lachapelle_disentanglement_2022}. Finally, we also developed a version of the architecture that included separate noise dimensions in the latent space of precipitation that are independent of ENSO and IOD. While this version of the architecture also converged well, we here decided to omit these additional dimensions. As they were not required for our studying the research questions asked in this paper, a simpler architecture was seen as preferable. \\

\begin{figure}[!htp]
\centering
  \noindent\includegraphics[width=18pc,angle=0]{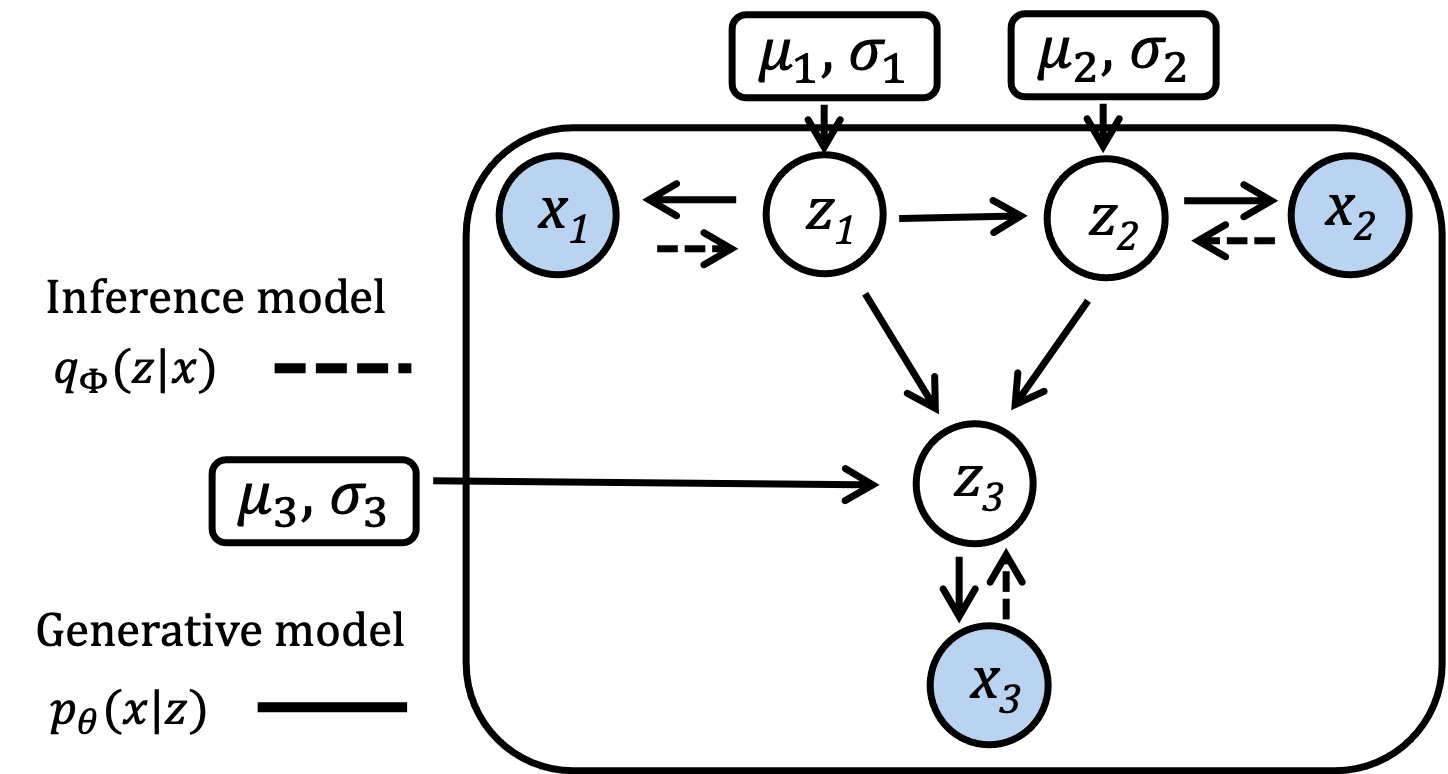}\\
  \caption{DAG-VAE architecture visualised as a probabilistic graphical model.}\label{architecture}
\end{figure}

To obtain the results presented in this paper, we trained the architecture visualised as a graphical model in Figure \ref{architecture}. Each encoder / decoder pair was implemented using three dense layers of decreasing dimensionality of 256, 128, and 64, respectively. A batch size of 128 and the ReLU activation function are chosen. For the implementation of the neural network architectures, the Python package keras \citep{chollet_keras_2015} was used. For a number of the evaluation metrics as well as the implementation of the two linear methods, the Python package scikit-learn \citep{pedregosa_scikit-learn_2011} was employed. The following weights were multiplied to the reconstruction loss terms: $\beta_{TP} = 0.2$, $\beta_{IO} = 0.3$ and $\beta_{PR} = 0.8$ to balance the difference in reconstruction loss resulting from the different area sizes. \\

For SEAS5, the model is trained 25 times for 100 epochs initialized with random weights with a learning rate of 0.001, and a lasso regularization parameter of 0.01 in the latent space, applying a train-validation-test split of 70\%-15\%-15\% and ensuring that data from one season is only part of one of the three datasets. For ERA5, the model is trained 25 times initialized from the weights of the trained SEAS model with a learning rate of 0.0005 and a lasso regularization parameter of 0.01 in the latent space, applying a train-test split of 80-20\%.


%



\clearpage


\bibliographystyle{plainnat}
\bibliography{references}

\end{document}